\documentclass[prx,twocolumn,10pt,aps,floatfix, footinbib,superscriptaddress]{revtex4-2}

\usepackage{graphicx, color, soul}
\usepackage{amssymb}
\usepackage[usenames,dvipsnames, pdftex]{xcolor}
\usepackage{ulem} 

\usepackage[mathscr]{eucal}

\usepackage[colorlinks, breaklinks, 
linkcolor=OrangeRed,
citecolor=RoyalBlue,
urlcolor=RoyalBlue]{hyperref}

\usepackage{multirow}

\usepackage[all]{hypcap} 
\newcommand{\Ac}{\mathscr{A}}

\newcommand{\Hc}{\mathscr{H}}

\newcommand{\eq}[1]{\begin{align}#1\end{align}}

\usepackage{physics}
\usepackage{calrsfs}

\begin{document}

\title{Out of equilibrium chiral higher order topological insulator on a $\pi$-flux square lattice}
\author{Ruchira V Bhat}
\affiliation{Department of Physics, Indian Institute of Technology Bombay, Mumbai 400076}
\author{Soumya Bera}
\affiliation{Department of Physics, Indian Institute of Technology Bombay, Mumbai 400076}

\date{\today}

\begin{abstract}
One of the hallmarks of bulk topology is the existence of robust boundary localized states. %
For instance, a conventional $d$ dimensional topological system hosts $d{-}1$ dimensional surface modes, which are protected by non-spatial symmetries.
Recently, this idea has been extended to higher order topological phases with boundary modes that are localized in lower dimensions such as in the corners or in one dimensional hinges of the system. 
In this work, we demonstrate that a higher order topological phase can be engineered in a nonequilibrium state when the time-independent model does not possess any symmetry protected topological states. 
The higher order topology is protected by an emerging chiral symmetry, which is generated through the Floquet driving. 
Using both the exact numerical method and an effective high-frequency Hamiltonian obtained from the Brillouin-Wigner perturbation theory, we verify the emerging topological phase on a $\pi$-flux square lattice.
We show that the localized corner modes in our model are robust against a chiral symmetry preserving perturbation and can be classified as `extrinsic' higher order topological phase. %
Finally, we identify a two dimensional topological invariant from the winding number of the corresponding sublattice symmetric one dimensional model. 
The latter model belongs to class AIII of ten-fold symmetry classification of topological matter. 
 \end{abstract}

\maketitle
\section{Introduction}
New states of matter can appear in nonequilibrium that may not have static counterparts.  
For instance, a discrete time crystal, as proposed by Wilczek~\cite{Timecrystals}, is a novel phase with spontaneously broken time translation symmetry, which is indeed found to be nonexistent in equilibrium~\cite{Masaki}. 
Similarly, through Floquet driving, new states of matter such as  topological states can be engineered, which may or may not have its static analogue~\cite{FloquetEngineering,FloquetTI_1,FloquetTI_3}.
In $d$ dimension the topological phases of matter are characterized by robust $d-n~( 0 < n \leq d)$ dimensional boundary localized modes, where $n{=}1$ represent a first order topological phase~\cite{asboth,book1,haldane,Kane,RMP1,RMP2,RMP3}.
The first order topological insulators~(TI) are usually classified by three non-spatial symmetries, such as the time reversal, the particle-hole, and the sublattice or the chiral symmetry~\cite{Kane,classification2,classification1}.
Generalization of such phases in the presence of time dependent Hamiltonian, which could be  modeled via periodic drive has been realized~\cite{Daniel1,Daniel2,Gloria,d1,d2,d3, d4,d5,d6,d7,d8,d11,FloquetEngineering,d10,FloquetTI_1,FloquetTI_2} and has also been investigated in various experiments subsequently~\cite{FloquetExp1,FloquetExp2,FloquetExp3}.

Recently, the notion of conventional TI's  is generalized to higher order topological insulators~(HOTI) with $n>1$~\cite{Taylor,hoti2,hoti3,SecondOrder_1,SecondOrder_2,Higherorder_1,Higherorder_3,Higherorder_4,Higherorder_5,Higherorder_6,Higherorder_7,Higherorder_8}.
The HOTI's are protected by the crystalline (spatial) symmetries such as the inversion, the mirror reflection, and the four fold rotation or space-time symmetries of both the bulk and the boundaries~\cite{Taylor,hoti1,hoti2,hoti3, Extrinsic1,Extrinsic2,chiral2,Higherorder_2}.
In the presence of these symmetries, HOTI phases appear as a consequence of the quantized higher electric multipole moments (quadrupole, octupole, etc.) of the bulk crystal~\cite{Taylor}.
If the HOTI phase is protected by the bulk gap and independent of the crystal termination, it is characterized as an intrinsic HOTI phase. 
On the contrary, when the non-trivial topology of the HOTI's are protected by both the bulk and the boundary gap it is recognized as an extrinsic HOTI phase~\cite{Extrinsic1,kane1}.

In equilibrium, various studies demonstrate that HOTI phases can be realized in the presence of non-spatial symmetries~\cite{Extrinsic2,Extrinsic1,chiral2}.
For instance, it has been observed that a two dimensional ($2$D) square lattice belonging to the BDI symmetry class hosts second order~($n{=}2$) topological states that are protected {\it only} by the chiral symmetry~\cite{chiral1}.
Further studies also put forth the method of construction of chiral symmetry protected $2$D as well as three dimensional ($3$D) extrinsic second order topological phases in lattice models in symmetry class AIII~\cite{chiral2}. 
The origin of this chiral symmetry protected equilibrium HOTI phases can be traced back to the corresponding non-trivial topology of the first order lower dimensional TI~\cite{highertolower,Secondorder_chiral1}.

In this work, we theoretically investigate the emergence of such a non-trivial chiral extrinsic dynamical HOTI phase via periodically (Floquet)  driving a trivial system.
Our model is a $\pi$-flux $2$D square lattice with arbitrary long-range hopping that belongs to the AI symmetry class.
Unlike the other works, where the Floquet higher order topological phases 
are protected by the space-time~\cite{spacetime,timeglide,Fhoti_spacetime1} or by the combination of both time reversal and four fold rotation~\cite{bitan1,bitan_adhip, Fhoti1,Seshadri} or by the antiunitary symmetries like particle-hole~\cite{bitan_particle,arjit_particle} and time reversal~ \cite{Floquet_superconductor} or by the mirror symmetries~\cite{Vincent, Vincent2,mirrorsymmetry1,Fhoti_mirror1}, in this work we show that a non-spatial unitary symmetry, such as the chiral symmetry, can also protect a dynamical extrinsic HOTI phase.
Our proposal involves modulation of the local site potential such that the sublattice symmetry is restored during the time evolution, which results in a chiral symmetry protected HOTI phase. 
To examine the HOTI phase we use Brillouin-Wigner (BW) perturbation theory~\cite{BW} to arrive at an effective time-independent Hamiltonian.
We show that the effective Hamiltonian is chiral symmetric, which belongs to the AIII symmetry class and possesses a second order topological phase that  reveals itself via multiple localized corner modes. 
The second order topology appearing in our model can be traced back to the chiral symmetry of the corresponding $1$D model and can be characterized by the $\mathbb{Z}$ topological invariant~\cite{chiral1, chiral2, kane1,Hayashi1,Hayashi2}.
Finally, by numerically studying the exact Floquet dynamics we confirm the validity of our high-frequency expansion and the existence of multiple HOTI phases with different topological numbers.
A similar study has been performed recently, where a 2D model has been constructed by stacking one dimensional ($1$D) equilibrium topological phases, and a possible chiral symmetric HOTI phase is shown to exist~\cite{Fhoti_chiral}. Unlike our proposal, in the stacked 1D model the periodic drive modulates the hopping amplitudes and not the local potential.

The remainder of the paper is organized as follows.
In Sec.~\ref{model} we introduce our periodically driven $\pi$-flux square lattice model.
The Floquet operator describing the exact dynamics of the model is discussed in Sec.~\ref{dynamics}.
In Sec.~\ref{highfrequency} we derive the time-independent effective Hamiltonian from the BW perturbation theory.
The significance of our driving protocol in restoring the sublattice symmetry is also discussed.
The band structure of sublattice symmetric time-independent effective Hamiltonian in reciprocal space is explained in Sec.~\ref{bulkhamiltonian}. 
We also explain how the Belancazar-Bernevig-Hughes~(BBH) model~\cite{Taylor,bitanantiunitary} appears in a certain limit of the effective model. 
In Sec.~\ref{symmetries} we analyze the symmetry properties of the corresponding effective Hamiltonian and show how chiral symmetry manifests in our model. 
Finally we discuss our numerical results in Sec.~\ref{results} and probe the stability of such extrinsic higher order topological phases in Sec.~\ref{stability}.  
%

\section{Model and methods}\label{model}
\begin{figure}[h]
\includegraphics[width=1.0\columnwidth]{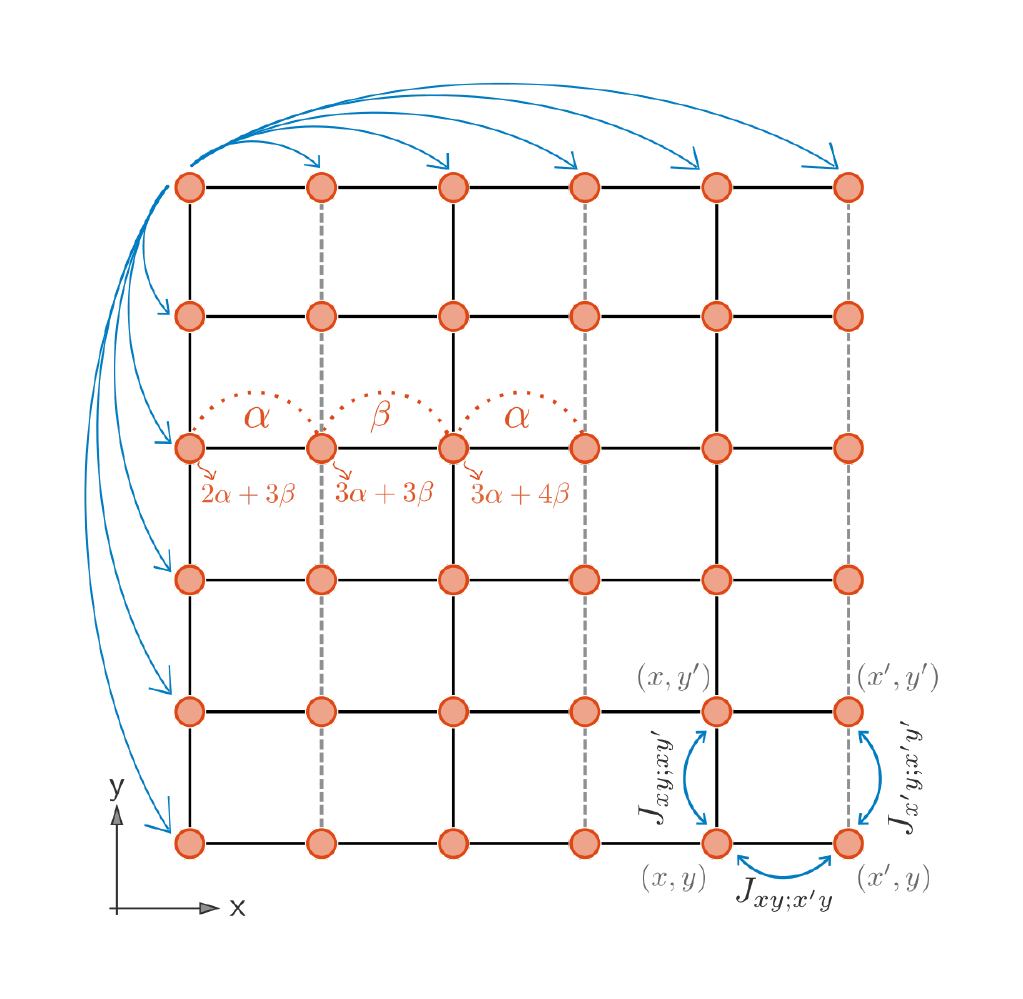}
\caption{Represents the $\pi$-flux threaded square lattice model. The model allows for long-range hopping as indicated, but no diagonal hopping are present. The time dependence appears on the onsite potential of each lattice sites. See Eq.~\eqref{modifiedrealqd} for further details. 
}\label{correctedlattice1}
\end{figure}

We consider a $2$D square lattice with  $\pi$- flux per plaquette as shown in Fig.~\ref{correctedlattice1}.
Previously, a similar model was studied in $d{=}1$ dimension to engineer first order topological phases from a trivial band structure~\cite{Gloria}.
%
The time-dependent Hamiltonian for spinless fermions on this lattice is given by,
\eq{\nonumber
\Hc(t) = \sum_{x',y'}\sum_{x,y} J_{xy;x'y'} c^{\dagger}_{x',y'} c_{x,y} + h.c \\ \label{modifiedrealqd} + \sum_{x,y} \Ac_{x,y} f(t) c^{\dagger}_{x,y} c_{x,y},}
where $x,y$ and $x',y'$  are the different lattice sites along which hopping are allowed,~ $J_{xy;x'y'}$ is the hopping amplitude between sites $(x,y)$ and $(x',y')$,~$c^{\dagger}_{x,y} ( c_{x,y})$ is the creation (annihilation) operator at the site $(x,y)$,~ $\Ac_{x,y}$ is the onsite potential at site $(x,y)$.
For even $x$, the hopping amplitude along $y$ direction takes an overall negative sign compared to the sites along $y$ for which $x$ is odd, i.e., $x = 2n (n = 1,2,3,...)$, $J_{2n,y;2n,y'} = -J_{2n-1,y;2n-1,y'}$  as shown by dotted lines in Fig.~\ref{correctedlattice1}.
While the hopping amplitude along $x$ direction $J_{xy;x'y}$ are the same for all $y$.

The onsite potential is,
\eq{\nonumber
\Ac_{2n} &= n(\alpha + \beta) ,\\ \nonumber
\Ac_{2n-1} &= n(\alpha + \beta) - \alpha ,}
where $n = 1,2,3,...$ takes integer values,~$2n$ and $2n-1$ represent lattice site either along $x$ or $y$, while $\alpha$, $\beta$ are the parameters that can be varied. 
The onsite potential on the lattice site $(x,y)$ is given by $\Ac_{x,y} = \Ac_{x} + \Ac_{y}$ where $\Ac_{x}= \Ac_{2n}~(\Ac_{2n-1}),~\Ac_{y} =\Ac_{2n}~(\Ac_{2n-1})$ for even (odd) $x$ and $y$.
Let us consider two neighbouring sites $x$ and $x+1$ with $x$ odd,~then $\Ac_{x+1,y} - \Ac_{x,y} = \alpha$ and if $x$ is even then $\Ac_{x+1,y} - \Ac_{x,y} = \beta$.
The same holds true for $y$ instead of $x$.
The onsite potential for few lattice sites and the corresponding dimer structure is shown in Fig.~\ref{correctedlattice1} with red dotted lines.
Note that the above pattern of $\Ac$ implies that $\Ac_{i,j} - \Ac_{i+1,j} = \Ac_{j,i} - \Ac_{j,i+1}$.

The periodic drive is a square pulse, which is represented as,
\eq{ \nonumber
f(t) = 
\begin{cases}
  \,-1 &\, \text{if} \quad 0 \leq t < \frac{T}{2}~, \\
  \,1  &\, \text{if} \quad \frac{T}{2} \leq  t < T,
\end{cases}}
where $T = \frac{2 \pi}{\omega}$ and $\omega$ is the frequency of the drive.
Without time dependence i.e. $f(t) =1$ for all time, the model belongs to the symmetry class AI, which does not have any symmetry protected first order topological phase.
Through Floquet engineering, we invoke sublattice symmetry in our $2$D system so that the corresponding $1$D model in class AIII have a chiral symmetry protected non-trivial first order topology~\cite{Gloria}.

\subsection{Exact dynamics using the Floquet operator}\label{dynamics}
We study the exact dynamics by diagonalizing the Floquet operator $U(T) = \mathscr{T}e^{-i \int_{0}^{T} dt ~\Hc(t)}$, where $\mathscr{T}$ is the time ordering operator.
Within the Floquet theory, periodically driven systems are described by the quasienergies and eigenstates of the Floquet operator.
In our model, the Floquet operator can be written as,
\eq{\label{floquetoperator}
U(T) &= e^{-i (\Hc_{\text{static}} + \Hc_{\text{drive}})\frac{T}{2}} e^{-i (\Hc_{\text{static}} - \Hc_{\text{drive}})\frac{T}{2}},} 
where,~$\Hc_{\text{static}}{=}\sum_{x',y'}\sum_{x,y} J_{xy;x'y'} c^{\dagger}_{x',y'} c_{x,y} + \text{h.c}$~and~$\Hc_{\text{drive}} {=} \sum_{x,y} \Ac_{x,y} f(t) c^{\dagger}_{x,y} c_{x,y}$.
The eigenvalue equation of the Floquet operator is $U(T)\ket{\psi} {=} e^{-i\epsilon{_\text{F}} T}\ket{\psi} {=} e^{-i\Hc_{\text{F}}T} \ket{\psi}$, where $\epsilon{_\text{F}}$ are the quasienergies and $\ket{\psi}$ are the Floquet eigenstates.
The effective Floquet Hamiltonian is therefore defined as,
\eq{
\label{effectivefloquetHamiltonian}\Hc_{\text{F}} = \frac{1}{-iT}\log{U(T)}.}
Finally the eigenvalues (quasienergies) of the effective Floquet Hamiltonian is calculated from the eigenvalue, $E$, of the Floquet operator $\epsilon{_\text{F}} {=} \frac{1}{-iT} \log (E)$.
To gain further insights into the emergent dynamical topological phase we analyze the high- frequency limit of the $\Hc_{\text{F}}$ as explained in the next section. 
\subsection{High-frequency approximation of the Floquet Hamiltonian}\label{highfrequency}
We here study the high-frequency regime ($J_{xy;x'y'} \ll \omega$) of the Floquet Hamiltonian~\eqref{effectivefloquetHamiltonian} using the BW perturbation theory~\cite{BW}.
As explained in Sec.~\ref{dynamics} the exact dynamics of a time periodic system can be studied using a time-independent effective Floquet Hamiltonian in an extended Hilbert space $\Hc \otimes \mathbb{T}$, where $\mathbb{T}$ comprises of the infinite dimensional space of states that represents the interaction between atoms and photons due to the drive~\cite{Eckardt}.
BW perturbation theory gives an effective time-independent Hamiltonian projected onto the finite dimensional zero photon subspace, which in the high-frequency regime gives us the energies and eigenstates of the effective Floquet Hamiltonian~\eqref{effectivefloquetHamiltonian}~\cite{BW}.
The significant contribution comes from the zeroth order and the first order terms.
Higher order terms are of the order $\mathit{O}(\frac{1}{\omega^{2}})$ and powers of $\omega$ greater than two.
Hence in the low drive period limit, we will neglect the higher order terms.
%

\begin{figure}[tbh]
\includegraphics[width=0.8\columnwidth]{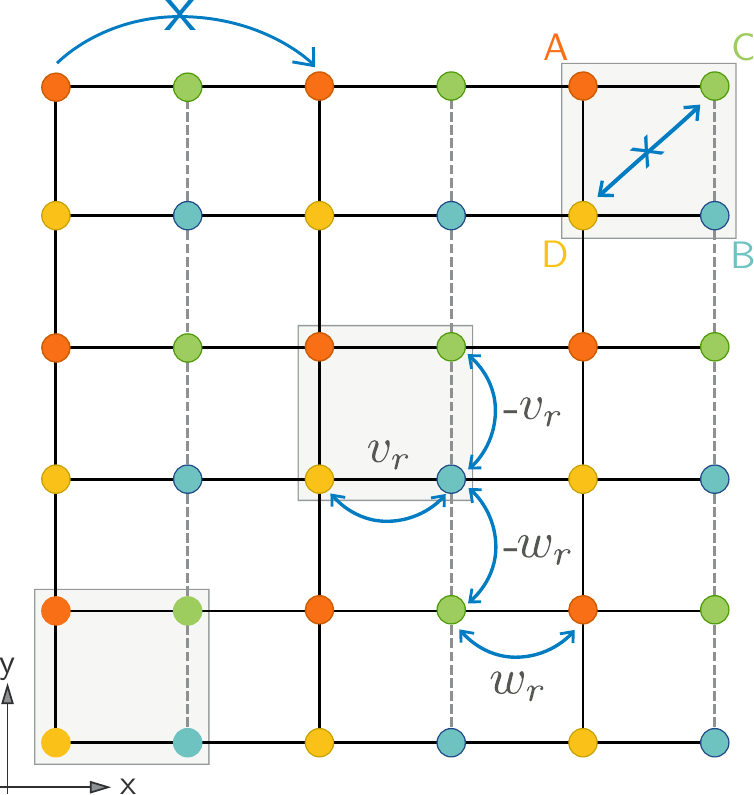}
\caption{The figure illustrates the effective square lattice after the high-frequency expansion using the BW perturbation theory~\eqref{realspacehopping}.
The unit cell is shown as the shaded region with four sublattices.
$v_{r}$ and $w_{r}$ represent the  hopping within the unit cell, and in between different unit cells respectively.
The long-range hopping between different sublattices is allowed but those within the same sublattice are discarded to restore the chiral symmetry.} \label{correctedlattice2}
\end{figure}

In order to proceed we first remove the time dependent onsite potential term in Eq.~\eqref{modifiedrealqd} by an unitary transformation $U {=} e^{-i\int \Ac_{x,y} f(t) c^{\dagger}_{x,y} c_{x,y}dt}$. 
The transformed Hamiltonian now contains hopping amplitude that are time dependent, $J_{xy;x'y'} \rightarrow J_{xy;x'y'} e^{i(\Ac_{x,y}-\Ac_{x',y'})\int f(t) ~dt}$.
Within the zeroth order expansion of BW perturbation theory, we get a time-independent effective Hamiltonian with renormalized time-independent hopping,  
\eq{ \label{renormalisedhopping}
J_{xy;x'y'}^{\text{R}} &= \frac{iJ_{xy;x'y'}\omega}{\pi(\Ac_{x,y}-\Ac_{x',y'})} \left(e^{-i\frac{(\Ac_{x,y}-\Ac_{x',y'})T}{2}}
 -1\right).}
where, $J_{xy;x'y} = r_0 e^{-\frac{|x-x'|}{\lambda}}$, $J_{xy;xy'} = r_{0} e^{-\frac{|y-y'|}{\lambda}}$, and $\lambda$ is the decay length, which we choose according to the range of hopping.
Due to $\pi$-flux per plaquette the following condition is obeyed $J_{2n,y;2n,y'}^{\text{R}} = - J_{2n-1,y;2n-1,y'}^{\text{R}}$.
In order to maintain the chiral symmetry, we tune the parameters in such a way that the hopping between odd-odd and even-even sites along each chain in $x$ and $y$ direction are suppressed.
Considering the odd-odd or even-even hopping between $(x,y)$ and $(x+2m,y)$ (or~$(x,y)$ and $(x,y+2m)$) with $m = \pm 1,2,3,...$ we get, $\Ac_{x+2m,y} - \Ac_{x,y} = \pm m(\alpha + \beta) = \Ac_{x,y+2m}-\Ac_{x,y}$.
For the effective odd-odd or even-even hopping, $J_{x,y;x+2m,y}^{\text{R}}$ and $J_{x,y;x,y+2m}^{\text{R}}$ to vanish we choose $ m(\alpha + \beta) = 2 \omega q$ where $q = 0,1,2,3,...$~.
These are the zeros of the function defined in Eq.~\eqref{renormalisedhopping} which are equally spaced in real space thus canceling out all long-range hopping amplitudes to the same sublattice.
Note such suppression of long-range hopping amplitude is not possible within a single frequency drive( such as a sine/cosine drive) because such drive protocol would renormalize the hopping to Bessel functions, where equally spaced zeros in real space is absent~\cite{10}.
Using a different approach i.e., the Floquet-Magnus expansion we verify that we recover similar physics as BW perturbation theory. 
The details of both the calculations are included in Appendix.~\ref{appendixa}.

\subsubsection{Time-independent effective bulk Hamiltonian} \label{bulkhamiltonian}
After eliminating the chiral symmetry breaking odd-odd and even-even hopping terms and calculating the other hopping amplitudes using Eq.~\eqref{renormalisedhopping} we get an equivalent two dimensional model with four atoms unit cell as shown in Fig.~\ref{correctedlattice2}.
The exact expressions for the chiral symmetry preserving hopping amplitudes along $x$ direction is,
\eq{\nonumber
J_{2n_{x},y;2n_{x}+r,y}^{\text{R}} &= w_{r} = \frac{-i r_{0} e^{-\frac{r}{\lambda}} \left(e^{i \pi\left[(r+1) q -\frac{\alpha}{\omega}\right]}-1\right)}{\pi\left[(r+1)q-\frac{\alpha}{\omega}\right]},\\ \label{realspacehopping}
J_{2n_{x}-r,y;2n_{x},y}^{\text{R}} &= v_{r} = \frac{-i r_{0}e^{-\frac{r}{\lambda}} \left(e^{i \pi \left[(r-1) q +\frac{\alpha}{\omega}\right]}-1\right)}{\pi\left[(r-1) q+\frac{\alpha}{\omega}\right]},}
where $n_{x}$, $n_{y}$ represents the unit cell number along $x$ and $y$ directions respectively, $r= |x-x'| = |y-y'|$ is the range of hopping and only odd values $r = 1,3,5,7,...$ are allowed.
$v_{r}$ is the forward hopping within the unit cell and $w_{r}$ is the forward hopping between two different unit cell.
As long as $r$ is the same in both directions and there are no diagonal hopping terms, the same expression applies for hopping in $y$ direction also.
Under periodic boundary conditions, the bulk Hamiltonian takes a $4 \times 4$ structure in the momentum space,
\eq{\label{matrixhamiltonian}
\hspace*{-1cm}\Hc(k_{x},k_{y}) = \sum_{r}
\begin{pmatrix}
0 & 0 & d_{r}(k_{x}) & d_{r}(k_{y})\\
 0& 0 & d^{*}_{r}(k_{y}) & d^{*}_{r}(k_{x}) \\
d^{*}_{r}(k_{x}) & d_{r}(k_{y})& 0& 0\\
d^{*}_{r}(k_{y})& d_{r}(k_{x}) & 0 & 0
\end{pmatrix}}
where $*$ denotes the complex conjugation and 
\eq{ \nonumber d_{r}(k_{x}) &= v_{r} e^{-ik_{x}\frac{(r-1)}{2}} + w^{*}_{r} e^{ik_{x}\frac{(r+1)}{2}},\\ \label{dvectors} 
d_{r}(k_{y}) &= v^{*}_{r} e^{-ik_{y}\frac{(r-1)}{2}} + w_{r} e^{ik_{y}\frac{(r+1)}{2}}.} 
The bulk Hamiltonian in Eq.~\eqref{matrixhamiltonian} can also be rewritten in terms of Pauli matrices as,
\eq{ \nonumber
\Hc(k_{x},k_{y}) = \sum_{r} \left[ d_{r}(k_{x}) (\Gamma_{1} + \Gamma_{2})\right] + h.c.  \\ \label{modifiedkspacepauliqd} +\left[ d_{r}(k_{y}) (\Gamma_{3} - \Gamma_{4}) \right] + h.c.,}
where,
\eq{ \nonumber \Gamma_{1} &= \left(\frac{\sigma_{x} + i \sigma_{y}}{2}\right) \otimes \left(\frac{\tau_{0}+ \tau_{z}}{2}\right),\\\nonumber
\Gamma_{2} &= \left(\frac{\sigma_{x} - i \sigma_{y}}{2}\right) \otimes \left(\frac{\tau_{0}- \tau_{z}}{2}\right),\\\nonumber
\Gamma_{3} &= \left(\frac{\sigma_{x} + i \sigma_{y}}{2}\right) \otimes \left(\frac{\tau_{x}+i \tau_{y}}{2}\right),\\ \nonumber
 \Gamma_{4} &=  \left(\frac{\sigma_{x} - i \sigma_{y}}{2}\right) \otimes \left(\frac{\tau_{x} + i \tau_{y}}{2}\right).}
$\sigma$ and $\tau$ are the Pauli matrices that represent the four unit cell sublattice structure.
The negative sign in Eq.~\eqref{modifiedkspacepauliqd} takes care of the $\pi$-flux in the square lattice (see Fig.~\ref{correctedlattice2}).
The edges of the effective square lattice model in Fig.~\ref{correctedlattice2} is similar to a $1$D topological dimer chain with inter-cell hopping amplitudes $w_{r}$ and intra-cell hopping amplitudes $v_{r}$.
If $v_{r}$~($w_{r}$) are equal along the $x$, $y$ directions then the edge Hamiltonian in all the four edges can be written as,
\eq{\label{edgeHamiltonian}
\Hc_{\text{edge}}(k) &= \left[d_{r}(k)\left(\frac{\sigma_{x} + i \sigma_{y}}{2}\right)\right] + \text{h.c.},}
where $k$ can be either $k_{x}$ or $k_{y}$ depending on the orientation of the edge. Here $d_{r}(k_{x})$ and $d_{r}(k_{y})$ are given in Eq.~\eqref{dvectors}.

In the limit where $v_{r} = v^{*}_{r}$,~$w_{r} = w^{*}_{r}$, and $r=1$, we recover the BBH Hamiltonian~\cite{bitanantiunitary,Taylor} from our Hamiltonian~\eqref{modifiedkspacepauliqd},
\eq{ \nonumber \Hc_{\text{BBH}} = (v_{r} + w_{r} \cos(k_{x}))\gamma_{4} + w_{r} \sin(k_{y}) \gamma_{3} +\\ \nonumber(v_{r} + w_{r} \cos(k_{y})) \gamma_{2} + w_{r}\sin(k_{y})\gamma_{1},}
where $\gamma$'s are the mutually anticommuting $4 \times 4$ Hermitian matrices that satisfy the Clifford algebra.
HOTI phase in the BBH model is characterized by the quantized quadrupolar moment of the bulk that possesses both the mirror reflection and the inversion symmetries~\cite{Taylor}.
%
%
\begin{figure*}[tbh]
\includegraphics[width=1.0\textwidth]{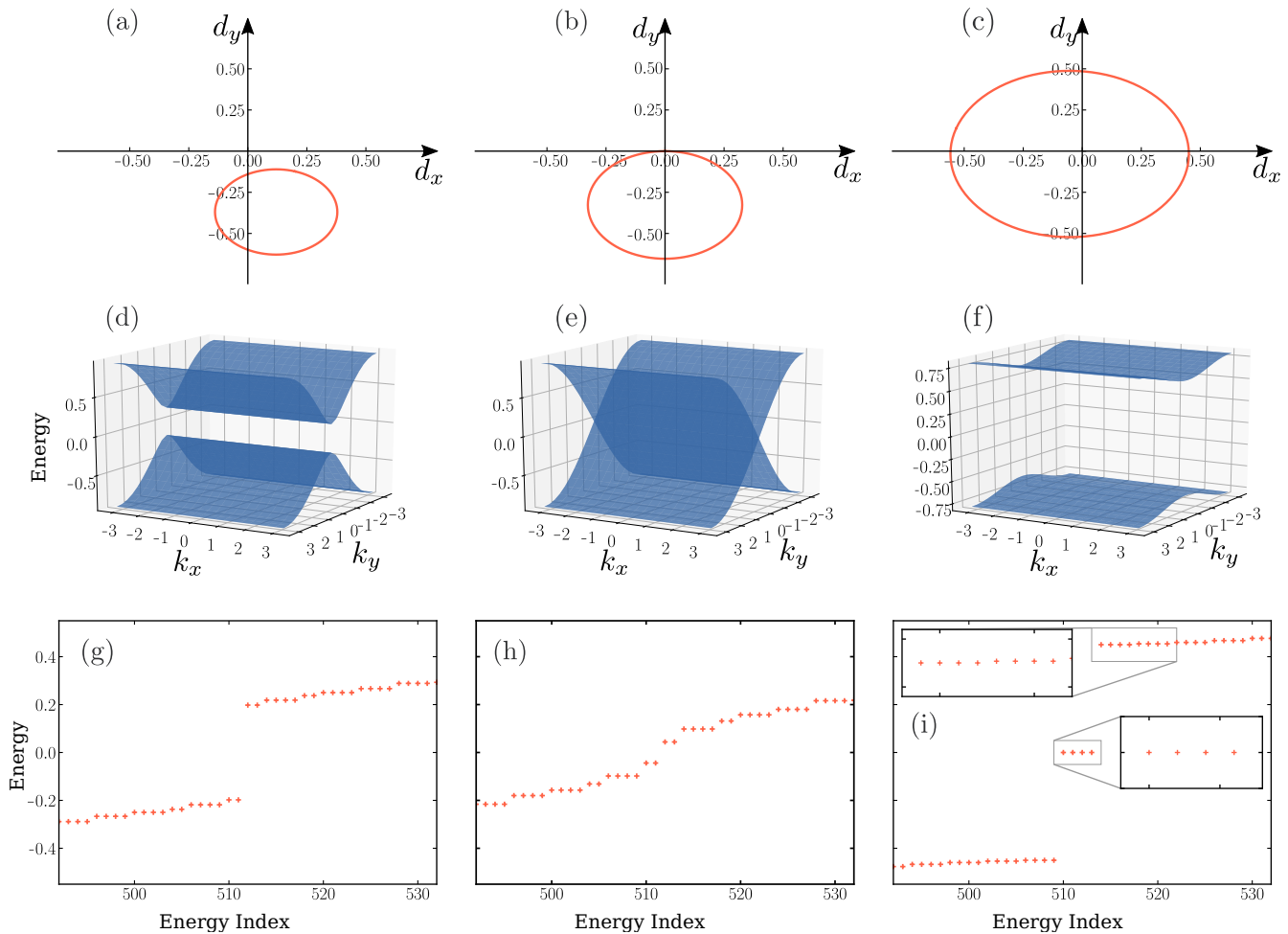}
 \caption{(a)-(c) Show the winding of $\vec{d}(k)$ as defined in $\Hc_{\text{edge}}(k)$~\eqref{edgeHamiltonian} (same along both the $x$ and $y$ direction) in the parameter space for three different values of $\alpha/\omega$.
 We consider only nearest neighbour hopping ($r=1$) and $\lambda {=} 1.5$ in all values of $\alpha/\omega$ with $r_{0}{=}q{=}1$.
(a) $\alpha/\omega {=} 0.8$, (b) $\alpha/\omega {=} 1$  (c) $\alpha/\omega {=} 1.8$ show the transition from a topologically trivial phase to a non-trivial phase in the corresponding one dimensional edge.
(d)-(f) The momentum space band structure of $\Hc (k_{x},k_{y})$~\eqref{modifiedkspacepauliqd} for similar parameters. 
The bands are doubly degenerate with a finite gap except at the transition point.
(g) -(h) Show the real space energy spectrum of the effective Hamiltonian calculated from the BW perturbation theory~\eqref{realspacehopping}.
The appearance of the zero energy state in (i) is correlated with the one dimensional topology as observed in (c). }
\label{result0}
\end{figure*}

\subsection{Symmetries of the effective bulk Hamiltonian}\label{symmetries}
In order to analyze the bulk Hamiltonian~\eqref{modifiedkspacepauliqd} we focus here on three non-spatial symmetries: time reversal ($\mathcal{T}$), particle-hole or charge-conjugation ($\mathcal{C}$), and sublattice or chiral symmetry ($\mathcal{S}$).
Here it is important to take into consideration the modified onsite potential that allow the restoration of the sublattice symmetry in the system as explained in Sec.~\ref{highfrequency}.

First, observe that there is no antiunitary operator that satisfies $U \Hc(k_{x},k_{y})U^{-1} = \Hc^{*}(-k_{x},-k_{y})$  or $\mathcal{T} \Hc(k_{x},k_{y})\mathcal{T}^{-1} = \Hc(-k_{x},-k_{y})$ for the bulk Hamiltonian~\eqref{modifiedkspacepauliqd}, where $\mathcal{T}$ represents the time reversal operation given by $\mathcal{T} = U K$, $U$ is a unitary operator and $K$ represents complex conjugation.
This happens because $v_{r}~(w_{r})~\neq~ v^{*}_{r}~(w^{*}_{r})$; hence the bulk Hamiltonian is not time reversal invariant.
Due to similar reason the particle-hole symmetry is also absent as there is no antiunitary operator in first quantised form that satisfies $\mathcal{C}\Hc(k_{x},k_{y})\mathcal{C}^{-1} = -\Hc(-k_{x},-k_{y})$ or $U \Hc(k_{x},k_{y})U^{-1} = -\Hc^{*}(-k_{x},-k_{y}) $ for Eq.~\eqref{modifiedkspacepauliqd} where $\mathcal{C}$ represents particle-hole operation given by $\mathcal{C} = U K $.
For the unitary operator $\mathcal{S} = \tau_{z} \otimes \sigma_{0}$, the bulk Hamiltonian satisfies $\mathcal{S}\Hc(k_{x},k_{y})\mathcal{S}^{-1} = -\Hc(k_{x},k_{y})$, which implies that the sublattice or chiral symmetry is preserved in the system~\cite{Adhip}.

Spatial symmetries that play a significant role in topological properties are, inversion, mirror reflection, and the four fold rotation symmetry.
Because of $v_{r}~(w_{r})~\neq~ v^{*}_{r}~(w^{*}_{r})$ our model does not have inversion and mirror reflection symmetries. 
Hence it turns out that the electric multipole moment, which commonly used to characterize an HOTI phase, is not quantized in our model~\cite{Taylor}. 

\section{Results: Bulk invariant and the HOTI phase} \label{results}
Within the ten-fold way of classification of topological insulators and superconductors~\cite{classification1, classification2}, the bulk Hamiltonian of our model belongs to the symmetry class AIII. 
In $2$D, class AIII does not have any topologically protected edge states as can be seen in the ten-fold classification schemes. 
However, the chiral symmetric dimer chain along the edges of the effective square lattice~\eqref{edgeHamiltonian} also belongs to the symmetry class AIII~\cite{classification1}; thus exhibits topological phase characterized by $\mathbb{Z}$ index, which is the one dimensional winding number $\nu$.
By tuning the hopping amplitude $J^{R}(r,\frac{\alpha}{\omega}, \lambda)$, higher $\nu$ is possible to obtain as shown previously in the context of one dimensional model by \textcite{Gloria}.
We find that it manifests itself in our model in 2D  by having multiple corner localized zero energy states. 
The edge Hamiltonian~\eqref{edgeHamiltonian} consists of two bands, and generically can be written as  $\Hc_{\text{edge}}(k) = \vec{d}(k)~.~\vec{\sigma}$, where $\sigma$'s are the Pauli matrices, with $\vec{\sigma} = \sigma_{x} \hat{x} + \sigma_{y} \hat{y} + \sigma_{z} \hat{z}$.
Since the $\Hc_{\text{edge}}$ belongs to class AIII,~$d_{z} = 0$.
Here the one dimensional winding number $\nu$ is defined in the first Brillouin zone as,
\eq{ \nonumber \nu &= \frac{1}{2\pi} \int_{0}^{2\pi} \left(\tilde{\vec{d}}(k) \times \frac{d}{dk} (\tilde{\vec{d}}(k))\right)_{z}~dk,}
where $\tilde{\vec{d}}(k)= \frac{\vec{d}(k)}{|\vec{d}(k)|}$.
\paragraph{Winding number and the band structure:} 
The $\vec{d}(k)$ of $\Hc_{\text{edge}}(k)$ is shown in the parameter space ($d_x, d_y$) in Fig.~\ref{result0}(a)-(c) at topologically trivial, transition point and non-trivial regions. 
Here we assume that $\Hc_{\text{edge}}(k_{x}) {=}\Hc_{\text{edge}}(k_{y})$, hence $\nu_{x}{=} \nu_{y}$ where $\nu_{x}$ and  $\nu_{y}$ are the winding  numbers along $x$ and $y$ direction of the effective square lattice in Fig.~\ref{correctedlattice2}.
Figure~\ref{result0}(a) represents a topologically trivial regime as the $\vec{d}(k)$ is not enclosing the origin and $\nu_{x} {=} \nu_{y} {=} 0$.
While in Fig.~\ref{result0}(b) $\nu_{x}$ and $\nu_{y}$ are not defined, since the $\vec{d}(k)$ passes through the origin.
$\vec{d}(k)$ encloses the origin once as sene in Fig.~\ref{result0}(c), hence $\nu_{x} {=} \nu_{y} {=} 1$. This corresponds to a topologically non-trivial phase.
We complement the above analysis of 1D $\Hc_{\text{edge}}$ with the exact diagonalization of the effective 2D Hamiltonian. 
Figure~\ref{result0}(d-f) show the band structure of $\Hc(k_{x},k_{y})$~\eqref{modifiedkspacepauliqd} and the corresponding real space energy spectrum with open boundary condition~\eqref{realspacehopping} for the same set of $\frac{\alpha}{\omega}$, $\lambda$, and $r$, see  Fig.~\ref{result0}(g-i). 
As observed with the closing of the band gap in 1D we also observe the closing of the 2D bulk gap, signifying the correspondence between the edge and the bulk spectrum.

The $\pi$-flux guarantees that for these specified values of $\frac{\alpha}{\omega}$ and $r$ in both the topological trivial and non-trivial regime, the four bands of $\Hc(k_{x},k_{y})$~\eqref{modifiedkspacepauliqd} are doubly degenerate with bulk gap as seen in Fig.~\ref{result0}(d,f). 
The band gap closes exactly at the phase transition point where the 1D edge also undergoes the topological transition (Fig. \ref{result0}(e)). 
This points us towards the existence of an extrinsic higher order topology in our model~\cite{Taylor}.
Finally, the appearance of zero energy corner states is in sync with the finite winding number in 1D and the bulk gap in the effective 2D spectrum, which indicates a topological phase, Fig.~\ref{result0}(i).
%
%
%

\begin{figure}[tbh]
\includegraphics[width=1.0\columnwidth]{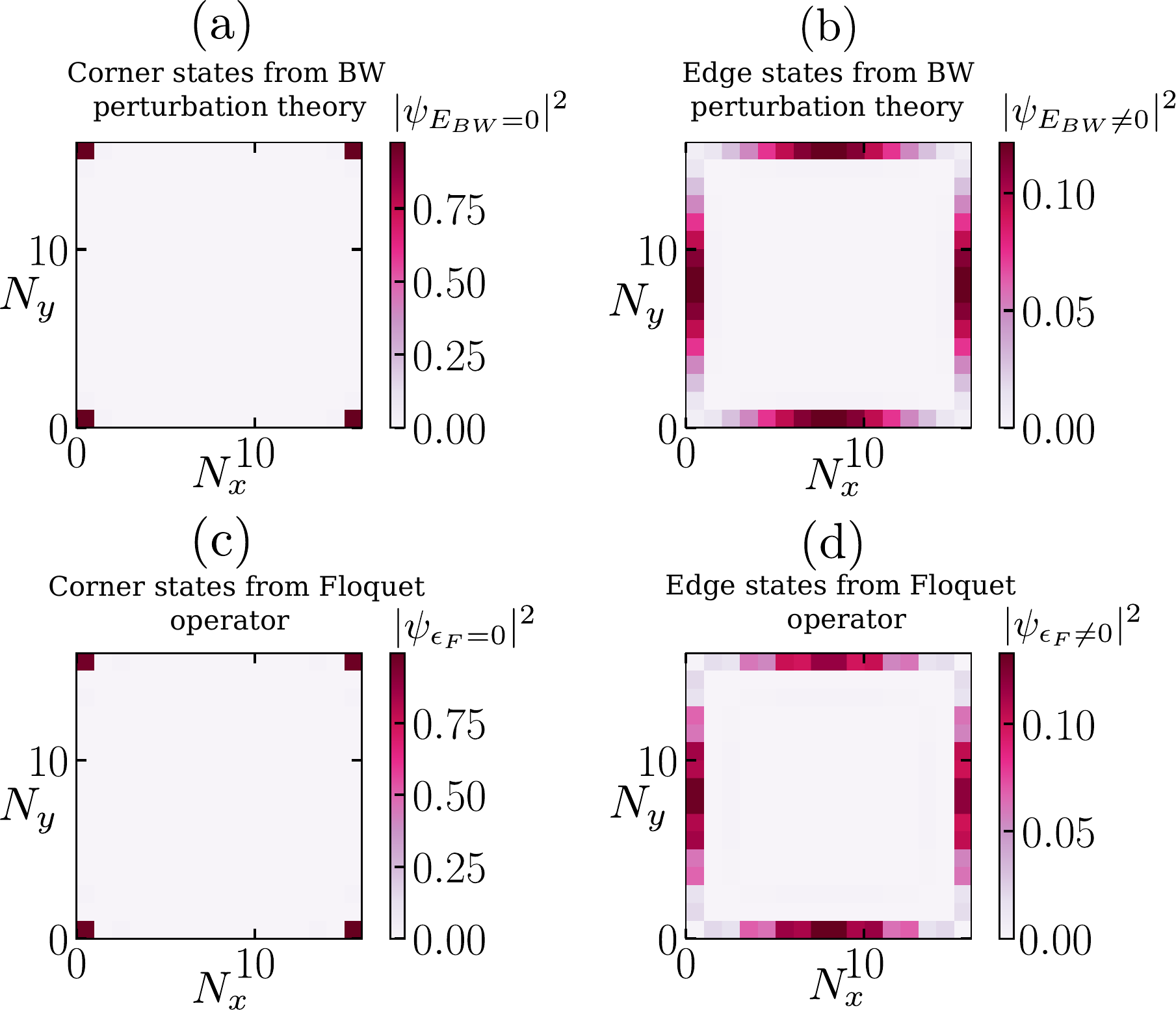}
\caption{(a) Shows the local density of states,  $|\psi_{E_{BW}{=}0}|^2$, for four zero energy corner states for a $16{\times} 16$ lattice with $\alpha/\omega {=} 1.8$,~$r{=}1$ and $\lambda {=} 1.5$.
(b) Represent the finite energy local density of states, $|\psi_{E_{BW}{\neq}0}|^2$,  of degenerate edge states  which are in the bulk as obtained from the effective high-frequency Hamiltonian~\eqref{realspacehopping}.
(c)-(d) Comparison of the states obtained from the exact diagonalization of the Floquet operator~\eqref{effectivefloquetHamiltonian}  in the high-frequency limit, $\omega{\approx}20 J_{xy; x^\prime y^\prime}$. }
\label{result1}
\end{figure}
\paragraph{Local density of states:} 
The local density of states for the four fold degenerate zero energy states and non-zero energy states obtained from the BW perturbation theory are shown in Fig.~\ref{result1}(a, b). 
As we see from Fig.~\ref{result1}(a,b), the zero energy states are localized at the corners and each set of non-zero energies are four fold degenerate that are localized at the edges of the square lattice.
The diagonalization of the Floquet operator~\eqref{effectivefloquetHamiltonian} for the same set of parameters also reproduces the expected corner states at zero quasienergy and four fold degenerate edge states as shown in Fig.~\ref{result1}(c,d).
This confirms the validity of the high-frequency approximation as derived in Sec.~\ref{highfrequency} in the regime $\omega > J_{xy;x'y'}$~\eqref{modifiedrealqd},  which although implicitly  depends on the parameters $r$ and $\lambda$.
The comparison of the energy spectrum of the effective Hamiltonian from the BW perturbation theory and the Floquet Hamiltonian from the exact dynamics is relegated to Appendix~\ref{appendixb}.
%

\begin{figure}[thb]
\includegraphics[width=1.0\columnwidth]{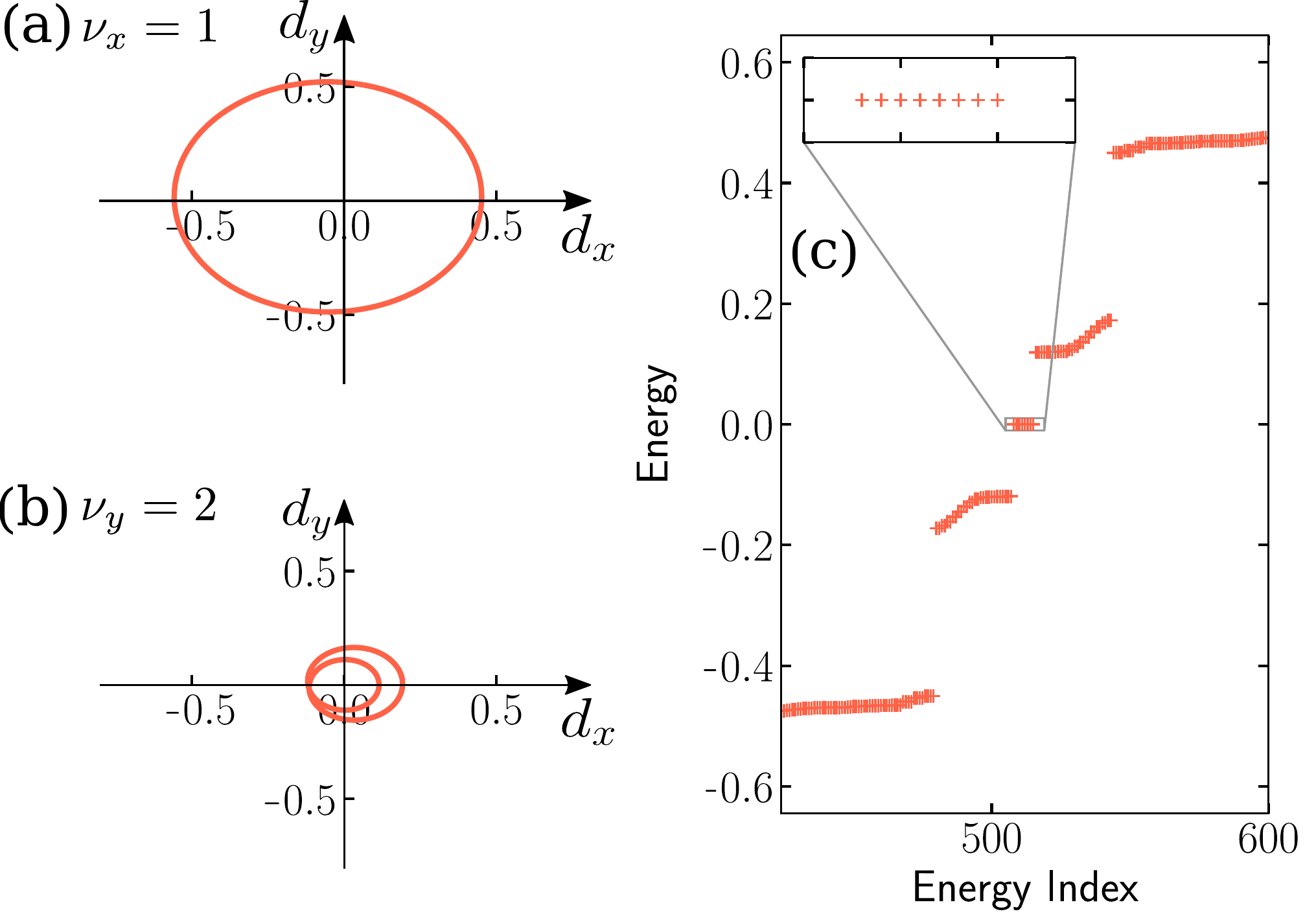}
\caption{(a) Shows the winding of $\vec{d}$ of $\Hc_{\text{edge}}(k)$ along the $x$ direction with $\nu_{x} {=} 1$, with $r{=}1, \alpha/\omega {=} 1.8$, and $\lambda {=} 1.5$. 
(b) Represents the winding along the $y$ direction with $\nu_{y} {=}2$ with $r{=}3, \alpha/\omega {=} 4.5$, and $\lambda {=} 1.5$.
(c) The $2$D energy spectrum of the effective time-independent Hamiltonian~\eqref{realspacehopping} confirms the existence of $8$ zero energy corner localized states, therefore,  $\nu_{2D} {=} 1 {\times} 2 {=} 2$.}\label{result2}
\end{figure}

\subsection{Two dimensional topological invariant from winding number}\label{invariant}
Since we get zero dimensional ($0$D) boundary states for sublattice symmetric $2$D system, our model belongs to the class of second order sublattice symmetry protected topological insulators.
From Fig.~\ref{result0} it is clear that the energy spectrum of  the bulk and the boundary gap close as we move from a trivial to a non-trivial phase, which points towards the extrinsic nature of the observed topological phase~\cite{Extrinsic1}.
The origin of these higher order topology is related to the $\mathbb{Z}$ topology of the corresponding one dimensional sublattice symmetric class AIII.
Similar sublattice symmetry protected higher order topological model is studied  in equilibrium in Refs.~[\onlinecite{Extrinsic1}], [\onlinecite{chiral1}] and  [\onlinecite{chiral2}].
Here the topological invariant is defined by the product of winding numbers of $\Hc_{\text{edge}}(k)$ in $x$ and $y$ direction \cite{chiral2}.
Each $\Hc_{\text{edge}}(k)$ is a sublattice symmetric chain belongs to class AIII with $\mathbb{Z}$ topological index similar to the 1D model~\cite{Gloria}.
Therefore the topological invariant in $2$D model is defined as,
\eq{\label{twodnu}
\nu_{2D} &= \nu_{x} \times  \nu_{y}, }
where $\nu_{x}$ and $\nu_{y}$ are the winding numbers of $\Hc_{\text{edge}}(k)$ in two directions.
The corner states appear when the winding number of two $\Hc_{\text{edge}}(k)$ intersecting at that corner is one.
For instance, in order to get four corner states we should have the winding number of all $\Hc_{\text{edge}}(k)$ to be one $\nu_{x} \times  \nu_{y} {=}1$.
Similarly for higher winding numbers of $\Hc_{\text{edge}}(k)$ gives us $4 \times \nu_{2D}$ corner localized sublattice symmetry protected zero energy states.
As mentioned previously that unlike first order topological insulators, here the protection of the HOTI phase is related to the non- trivial topology of the $\Hc_{\text{edge}}(k)$. 
As an illustration in Fig.~\ref{result2}(a,b) we show a situation where the one dimensional winding numbers are $\nu_{x} {=} 1$ and $\nu_{y} {=} 2$ for $\Hc_{\text{edge}}(k)$.
As expected we observe the appearance of $8$ corner states in the bulk spectrum, see Fig.~\ref{result2}(c), confirming the bulk invariant~\eqref{twodnu}. 
Generalizing this idea we get higher values for $\nu_{2D}$ and multiple sublattice symmetry protected corner states corresponding to the higher winding numbers in one dimension, which only depends on the range of the hopping.

\begin{figure}[h]
\includegraphics[width=0.9\columnwidth]{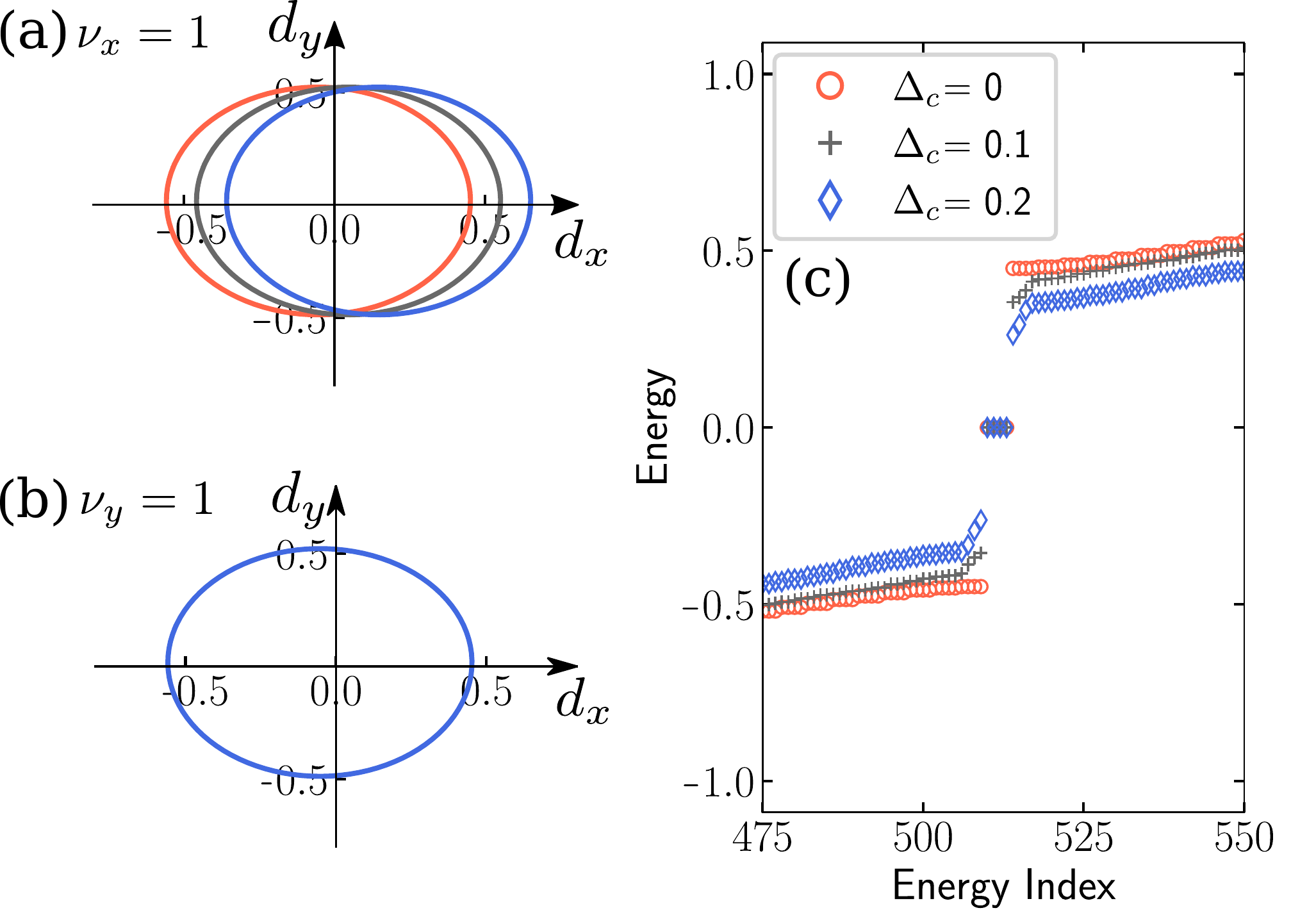}
\caption{Probe the stability of the corner state in the presence of a perturbation that preserves the chiral symmetry of $\Hc(k_{x},k_{y})$~ \eqref{modifiedkspacepauliqd}. 
The perturbation acts in the $x$-direction as observed by the change in $\vec{d}(k_x)$.
(a)-(b) Show the winding numbers $\nu_{x}$ and $\nu_{y}$ respectively, for perturbation $\Delta_{c}{=}\{0, 0.1, 0.2\}$.
(c) The corresponding $2$D energy spectrum shows the zero energy states for values of $\Delta_{c}$ that is smaller than the bulk gap.}\label{chiral1}
\end{figure}
\begin{figure}[h]
\includegraphics[width=0.9\columnwidth]{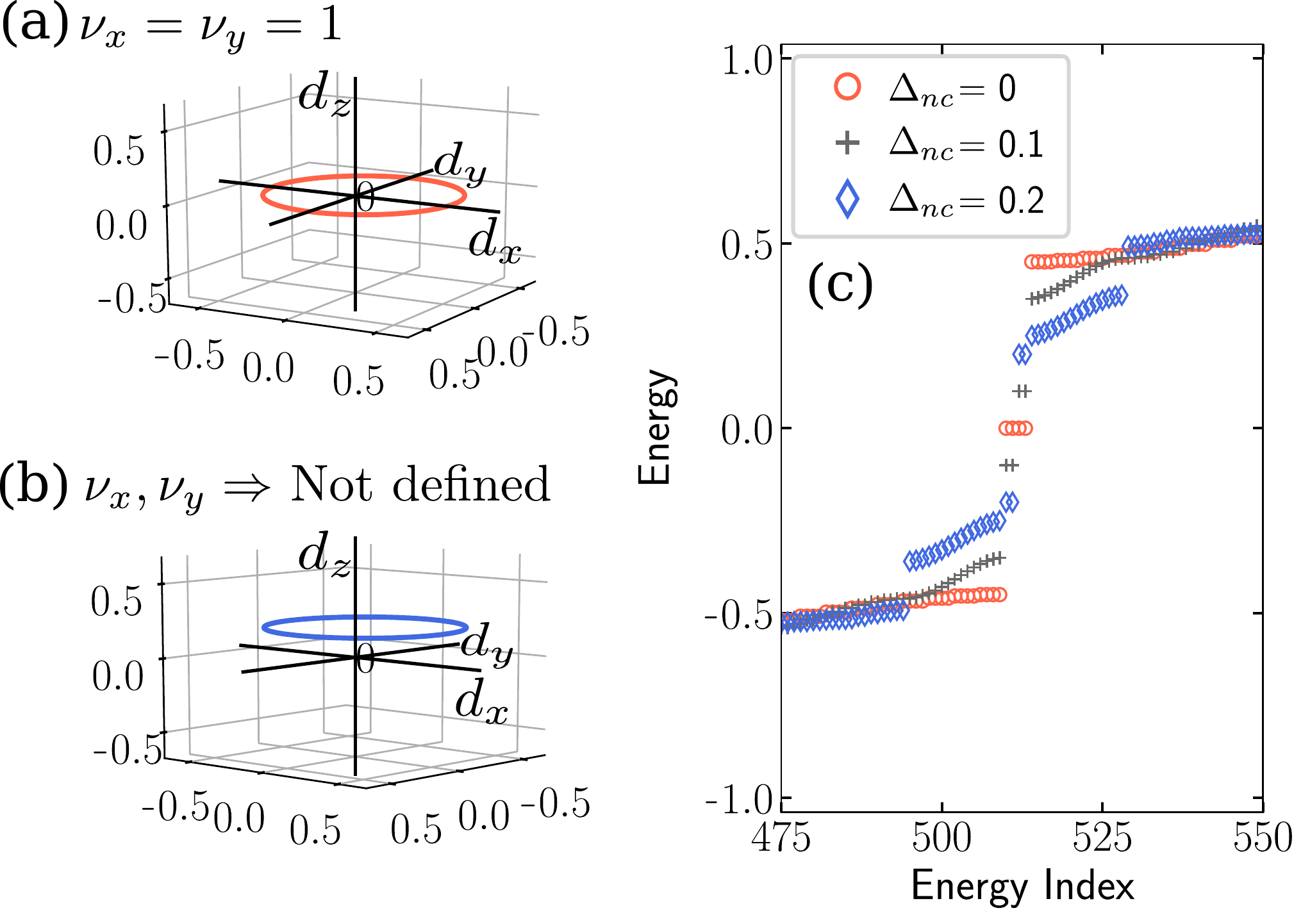}
\caption{(a)-(b) The winding of $\vec{d}(k)$ for $\Delta_{nc} {=} {0,0.2}$.
The perturbation introduces a non-zero mass term in the $z$-direction, thus breaking the chiral symmetry of the model. 
(c) Shows the corresponding $2$D energy spectrum, where we observe the disappearance of the zero energy states for any finite $\Delta_{nc}$.}\label{chiral2}
\end{figure}

\subsection{Stability of the corner state under perturbation}\label{stability}
\paragraph{Chiral symmetric perturbation:} 
The stability of the corner states can be traced back to the presence of the chiral symmetry of the model~\eqref{matrixhamiltonian}. 
This can be further confirmed by adding a chiral symmetry breaking and chiral symmetry preserving perturbation to the bulk Hamiltonian and monitoring the response of the HOTI phase through the edge modes. 
For instance, if we add a generic perturbation of the form $\Delta_{c}(\tau_{x} \otimes \sigma_{0})$ to the bulk $\Hc(k_x, k_y)$~\eqref{matrixhamiltonian} it does not break the chiral symmetry as $\mathcal{S}\Delta_{c}(\tau_{x} \otimes \sigma_{0}) \mathcal{S}^{-1} {=} -\Delta_{c}(\tau_{x} \otimes \sigma_{0})$, where $\Delta_c$ is the strength of the perturbation. 
This term results to a modified hopping along the $x$ direction in the square lattice.
Hence the $\Hc_{\text{edge}}(k)$ along $x$ gets modified, while the $\Hc_{\text{edge}}(k)$ along $y$ direction remains unchanged.
Figure~\ref{chiral1} shows the corresponding $\vec{d}(k)$ along $x$ and $y$ direction for the perturbed chiral symmetric $\Hc_{\text{edge}}(k)$~\eqref{edgeHamiltonian} and the corresponding $2$D energy spectrum of the effective time-independent Hamiltonian~\eqref{realspacehopping} for $\alpha/\omega = 1.8$, $\lambda {=} 1.5$ and $r {=} 1$. 
As observed the zero energy states and the winding numbers $\nu_{x}{=}1, \ \nu_{y}{=}1$ are preserved as long as the strength of the perturbation $\Delta_{c}$ is less than the energy gap of the original model.
Also since the perturbation is only along the $x$ direction hopping, the winding number $\nu_{y}$ remains invariant, while $\nu_x$ seems to shift with increasing the strength of the perturbation $\Delta_{c}$. 

\paragraph{Chiral symmetry breaking perturbation:} 
A perturbation of the form $\Delta_{nc}(\tau_{z} \otimes \sigma_{z})$ to the bulk Hamiltonian breaks the chiral symmetry as $\mathcal{S}\Delta_{nc}(\tau_{z} \otimes \sigma_{z})\mathcal{S}^{-1} \neq - \Delta_{nc}(\tau_{z} \otimes \sigma_{z})$, with $\Delta_{nc}$ being the strength of the perturbation.
This term is equivalent to adding a staggered onsite potential to the sublattices along each of the edges.
Therefore, we have $d_{z}(k) {=} \Delta_{nc}$ for the $\Hc_{\text{edge}}(k)$ and as $d_{z}(k)$ is independent of $k$, it shifts the vector $d_{x}(k)~ \hat{x} + d_{y}(k)~ \hat{y}$ in the $(x,y)$ plane along the $z$ direction, see Fig.~\ref{chiral2}(a, b).
The non zero winding number is observed when $\Delta_{nc} {=} 0$ with $\alpha/\omega {=} 1.8$, $\lambda {=} 1.5$, $r = 1$.
For finite $\Delta_{nc} {=} 0.05$ with similar parameters, sublattice symmetry is broken in both the direction.
The resulting $\Hc_{\text{edge}}(k)$ with no non-spatial symmetries belongs to class A that does not have any topological invariant in one dimension~\cite{classification1}.
Correspondingly in $2$D for any finite values of $\Delta_{nc}$ (but still much smaller than the bulk gap), the corner states are lifted away from the zero energy as seen in Fig.~\ref{chiral2}(c). 
This further confirms our assertion that the zero energy corner states of the HOTI phase are protected by the sublattice symmetry in this model.

\section{Conclusion}\label{conclusions}
In this work, we show that via periodic driving it is possible to engineer extrinsic HOTI phases that are protected by a non-spatial symmetry such as the chiral symmetry. 
We consider a model of square lattice with $\pi$-flux, which is driven to modulate the local potential in each drive cycle. 
In the high-frequency regime, we derive an effective model using Brillouin-Wigner perturbation theory, and confirm that the effective model hosts a second order topological phase with corner localized zero energy state. 
The high-frequency approximation is substantiated by the exact calculation of the Floquet operator in an appropriate limit. 
The $\mathbb{Z}$ topological invariant, defining these corner states are obtained from the one dimensional topological invariant (winding number) of the corresponding $\Hc_{\text{edge}}(k)$.
Finally, we show that by increasing the range of hopping the number of corner states can be achieved within our drive protocol and is reflected via higher topological invariant $\nu_{2D}$. 
Our proposal of 2D HOTI phase can  be realized  in trapped ion experiments~\cite{trappedion} as local modulation of ions and long-range hopping is natural in these systems.  Finally recently discovered highly controllable quantum dot arrays would also be a suitable candidate to realize the proposed HOTI phase~\cite{Quantumdot1,Quantumdot2,Quantumdot3,Quantumdot4}.

The static $2$D model can be further extended to three dimensions with $\pi$-flux threaded through the opposite faces of a cubic unit cell.
Here we anticipate to find a third order chiral symmetry protected extrinsic HOTI phase with zero energy corner states within the same driving protocol.
The number of corner states, therefore, can be obtained from the winding number,  which will be the product of winding numbers of three $\Hc_{\text{edge}}(k)$ intersecting at a given corner~\cite{kane1}. 
Such a study of the 3D model is currently underway. 
Finally, the fate of the HOTI phase in the presence of electron-electron interaction would be an interesting future direction to explore within the Floquet perturbation theory~\cite{Floquetperturbation1,Floquetperturbation2}.

\section{ACKNOWLEDGMENTS}
We would like to thank S. Pujari, B. Roy, and A. Saha for several insightful discussions. We also thank B. Roy and A. Saha for a critical reading of the manuscript. SB acknowledges support from Department of Science and Technology~(DST), India, through Ramanujan Fellowship Grant No. SB/S2/RJN-128/2016, Early Career Award No. ECR/2018/000876, Matrics No. MTR/2019/000566, and MPG for funding through the Max Planck Partner Group at IITB. RB would like to thank DST-INSPIRE fellowship No.~IF180067 for funding.

\bibliography{reference}

\appendix
\begin{widetext}
\section{Brillouin-Wigner perturbation theory and Floquet-Magnus expansion}\label{appendixa}
The exact Hamiltonian of our model is,
\eq{ \nonumber
\Hc(t) = \sum_{y}\sum_{x,x'} J_{xy;x'y} c^{\dagger}_{x',y} c_{x,y} +~h.c ~~+ ~\sum_{x = 1,3,5,...}\sum_{y,y'} J_{xy';xy} c^{\dagger}_{x,y'} c_{x,y} +~h.c~~-~ \sum_{x = 2,4,6,..} \sum_{y,y'} J_{xy';xy} c^{\dagger}_{x,y'} c_{x,y} + ~h.c\\ \label{Appendix2dexact} + \sum_{x,y} \Ac_{x,y} f(t) c^{\dagger}_{x,y} c_{x,y},}
The creation and annihilation operators obey the anticommutation relation given by $\left\lbrace c_{x,y},c^{\dagger}_{x',y'}\right\rbrace {=} \delta_{x.x'}\delta_{y,y'}$.
 Let the $\Hc_{\text{drive}}(t) =\sum_{x,y} \Ac_{x,y}f(t)c^{\dagger}_{x,y} c_{x,y}$ , $U(t) = e^{-i \int dt \Hc_{\text{drive}}(t)}$  and $F(t) = \int f(t) dt$.
Following same procedure as in Ref.~\cite{Gloria} here also we make a rotating frame transformation,
\eq{\label{appendixrotating}
\tilde{\Hc}(t) &= U^{\dagger}(t) \Hc(t) U(t) - i U^{\dagger}\partial_{t}U(t)
}
We solve for each term which is obtained by substituting Eq.~\eqref{Appendix2dexact} in Eq.~\eqref{appendixrotating}.~Taking the first term,
\eq{\nonumber
e^{iF\Ac_{x,y}c^{\dagger}_{x,y}c_{x,y}}\left(J_{x'y';x''y'}c^{\dagger}_{x',y'}c_{x'',y'}\right)e^{-iF\Ac_{x,y}c^{\dagger}_{x,y}c_{x,y}} &= J_{x'y';x''y'}c^{\dagger}_{x',y'}c_{x'',y'} + J_{x'y';x''y'}i F(t) \left[\Ac_{x,y}c^{\dagger}_{x,y}c_{x,y},c^{\dagger}_{x',y'}c_{x'',y'}\right] + ...,\\ \label{commutaion1}
&= J_{x'y';x''y'}c^{\dagger}_{x',y'}c_{x'',y'} + J_{x'y';x''y'}iF(t) \left(\Ac_{x',y'} - \Ac_{x'',y'}\right)c^{\dagger}_{x',y'}c_{x'',y'}
}
The commutation relation can be simplified in the following way to arrive at the final form of Eq.~\eqref{commutaion1},
\eq{\nonumber
\Ac_{x,y}\left[c^{\dagger}_{x,y}c_{x,y}, c^{\dagger}_{x',y'}c_{x'',y''}\right] &= \Ac_{x,y}\left( c^{\dagger}_{x,y}\left[c_{x,y},c^{\dagger}_{x',y'}c_{x'',y''}\right]  +\left[c^{\dagger}_{x,y}, c^{\dagger}_{x',y'}c_{x'',y''}\right] c_{x,y}\right)
}
\eq{\nonumber
= \Ac_{x,y}\left(c^{\dagger}_{x,y}\left\lbrace c_{x,y},c^{\dagger}_{x',y'}\right\rbrace  c_{x'',y''} - c^{\dagger}_{x,y}c^{\dagger}_{x',y'}\left\lbrace c_{x,y},c_{x'',y''}\right\rbrace +
 \left\lbrace c^{\dagger}_{x,y},c^{\dagger}_{x',y'}\right\rbrace c_{x'',y''}c_{x,y} - c^{\dagger}_{x',y'}\left\lbrace c^{\dagger}_{x,y},c_{x'',y''}\right\rbrace c_{x,y}\right),\\ \nonumber
 = \Ac_{x,y} \left( c^{\dagger}_{x,y} \delta_{xx'}\delta_{yy'}c_{x'',y'} - c^{\dagger}_{x',y'} \delta_{xx''}\delta_{yy'}c_{x,y}\right),\\ \nonumber
 = \Ac_{x',y'}c^{\dagger}_{x',y'} c_{x'',y'} - A_{x'',y'}c^{\dagger}_{x',y'} c_{x'',y'}
}
Similarly solving the second term obtained by substituting $\Hc(t)$ \eqref{Appendix2dexact} in  Eq.~\eqref{appendixrotating} we get,
\eq{\label{appendixsecond}
e^{iF\Ac_{x,y}c^{\dagger}_{x,y}c_{x,y}}\left(J_{x'y';x'y''}c^{\dagger}_{x',y'}c_{x',y''}\right)e^{-iF\Ac_{x,y}c^{\dagger}_{x,y}c_{x,y}} &= J_{x'y';x'y''}c^{\dagger}_{x',y'}c_{x',y''} + J_{x'y';x'y''}iF(t) \left(\Ac_{x',y'} - \Ac_{x',y''}\right)c^{\dagger}_{x',y'}c_{x',y''}
}
The third term in $\Hc(t)$ substituted Eq.~\eqref{appendixrotating} takes the same value as that of Eq.~\eqref{appendixsecond} with an overall negative sign and
the fourth term gets cancelled due to $-i U^{\dagger}\partial_{t}U(t)$.
Hence the Hamiltonian in rotating frame can be written as,
\eq{\nonumber
\tilde{\Hc}(t) = \sum_{y} \sum_{x,x'} J_{xy;x'y} c^{\dagger}_{x,y}c_{x',y} e^{iF(t)(\Ac_{x,y} - \Ac_{x'y})} + \sum_{x=1,3,5,..} \sum_{y,y'} J_{xy;xy'}c^{\dagger}_{x,y}c_{x,y'} e^{iF(t)(\Ac_{x,y} - \Ac_{x,y'})}  \\ \label{Appendixrotating}- \sum_{x=2,4,6,..} \sum_{y,y'} J_{xy;xy'}c^{\dagger}_{x,y}c_{x,y'} e^{iF(t)(\Ac_{x,y} - \Ac_{x,y'})}
}
The zeroth order term in the BW perturbation theory as well as Floquet-Magnus expansion is given by \cite{Eckardt,BW},
\eq{ \nonumber
\Hc^{0}_{BW} &= \frac{\int_{0}^{T} dt_{1} \tilde{\Hc}(t_{1})}{T},
}
\eq{ \nonumber
\Hc^{0}_{BW} =  \sum_{y} \sum_{x,x'}\frac{iJ_{xy;x'y}\omega}{\pi(\Ac_{xy}-\Ac_{x'y})}\left(e^{-i\frac{(\Ac_{xy}-\Ac_{x'y})T}{2}} -1\right) c^{\dagger}_{x,y}c_{x',y} + \sum_{x=1,3,5,..} \sum_{y,y'}\frac{iJ_{xy;xy'}\omega}{\pi(\Ac_{xy}-\Ac_{xy'})}\left(e^{-i\frac{(\Ac_{xy}-\Ac_{xy'})T}{2}} -1\right) c^{\dagger}_{x,y}c_{x,y'} \\ \label{appendixfloquet2dqd} -\sum_{x= 2,4,6,..} \sum_{y,y'}\frac{iJ_{xy;xy'}\omega}{\pi(\Ac_{xy}-\Ac_{xy'})}\left(e^{-i\frac{(\Ac_{xy}-\Ac_{xy'})T}{2}} -1\right) c^{\dagger}_{x,y}c_{x,y'}}
where $x,x'$ and $y,y'$ represents lattice sites in the $x$ and $y$ direction respectively.

The first order term in BW expansion for a general time dependent model  $H(t) = \sum_{i,j} J_{i,j}(t) c^{\dagger}_{i} c_{j}$ with only the nearest neighbour hopping is,
\eq{\label{firstorderbw}
H_{BW}^{1} &= \frac{1}{n\omega} \sum_{m \neq 0} \mathit{H}_{0,n} \mathit{H}_{n,0} = \frac{1}{n\omega} \sum_{m \neq 0} J_{i,k}^{-n} J_{k,j}^{n} c^{\dagger}_{i} c_{j} 
}
where,~$\mathit{H}_{0,n}$ and $ \mathit{H}_{n,0}$ are the fourier components of time dependent original Hamiltonian \cite{BW} and the fourier component of hopping amplitude is  $J_{i,j}^{m-n} = \frac{1}{T}\int_{0}^{T} dt~J_{i,j}(t) e^{i(m-n)\omega t}$.
Using this the first order term for Eq.~\eqref{Appendixrotating} is calculated.
Taking the first term obtained by using Eq.~\eqref{firstorderbw} for Eq.~\eqref{Appendixrotating},
\eq{\nonumber
J_{xy;x'y}^{-n} &= \frac{1}{T}\int_{0}^{T} dt~ J_{xy;x'y} e^{-i n \omega t} e^{i F (t) (\Ac_{x,y} - \Ac_{x',y})} dt,\\ \nonumber
&= \frac{J_{xy;x'y}}{T}\left( \int_{0}^{\frac{T}{2}} dt~e^{-i n \omega t} e^{-i t (\Ac_{x,y} - \Ac_{x',y})} + \int_{\frac{T}{2}}^{T} dt~e^{-i n \omega t} e^{i(\Ac_{x,y} - \Ac_{x',y})(-T+t)} \right),\\ \label{firsthopping}
&= \frac{i J_{xy;x'y}\omega}{2\pi}\left((-1)^{n} e^{-i(\Ac_{x,y} - \Ac_{x',y})\frac{T}{2}} -1 \right)\left(\frac{1}{(\Ac_{x,y} - \Ac_{x',y})+n \omega} + \frac{1}{(\Ac_{x,y} - \Ac_{x',y})-n \omega}\right)
}
Similarly the integration of  $J_{x'y;x''y}^{n}$ gives,
\eq{\label{second hopping}
J_{x'y;x''y}^{n} &= \frac{i J_{x'y;x''y}\omega}{2\pi}\left((-1)^{n} e^{-i(\Ac_{x',y} - \Ac_{x'',y})\frac{T}{2}} -1 \right)\left(\frac{1}{(\Ac_{x',y} - \Ac_{x'',y})-n \omega} + \frac{1}{(\Ac_{x',y} - \Ac_{x'',y})+n \omega}\right)
}
The same equation applies for second and third terms in Eq.~\eqref{Appendixrotating} with change along $y$. 
If we consider only nearest neighbour hopping in the original Hamiltonian then two paths can be considered for the $2$D square lattice that satisfy Eq.~\eqref{firstorderbw} with a common lattice point in between. 
The first path is  $(x,y) \longrightarrow (x+1,y) \longrightarrow (x+2,y)$.
The second path from $(x,y) \longrightarrow (x+1,y) \longrightarrow (x+1,y+1)$ is diagonal.
From Eq.~\eqref{firstorderbw} and using Eq.~\eqref{firsthopping} and Eq.~\eqref{second hopping}, the first order term corresponding to the first term in Eq.~\eqref{Appendixrotating} is,
\eq{ \nonumber
\Hc_{BW}^{1} &= \sum_{n \neq 0} \frac{1}{n \omega} J_{xy;x'y}^{-n} J_{x'y;x''y}^{n} c^{\dagger}_{x,y} c_{x'',y}
}
where,
\eq{\label{productfirstorder}
J_{xy;x'y}^{-n} J_{x'y;x''y}^{n}  &= \frac{- J_{xy;x'y}J_{x'y;x''y} \omega^{2}}{4\pi^{2}}(a \times b)\left((-1)^{n} e^{-i(\Ac_{x,y} - \Ac_{x',y})\frac{T}{2}} -1 \right)\left((-1)^{n} e^{-i(\Ac_{x',y} - \Ac_{x'',y})\frac{T}{2}} -1 \right)
}
where,
\eq{\nonumber
a &= \left(\frac{1}{(\Ac_{x,y} - \Ac_{x',y})+n \omega} + \frac{1}{(\Ac_{x,y} - \Ac_{x',y})-n \omega}\right),\\ \nonumber
b &= \left(\frac{1}{(\Ac_{x',y} - \Ac_{x'',y})-n \omega} + \frac{1}{(\Ac_{x',y} - \Ac_{x'',y})+n \omega}\right)
}
Along the two paths mentioned above the hopping breaks the chiral symmetry of the system.
To maintain chiral symmetry, we should tune these second neighbour hopping to zero. 
For the first path, $\Ac_{x,y} - \Ac_{x+1,y} = \alpha$ and $\Ac_{x+1,y} - \Ac_{x+2,y} = \beta$ for odd $x$. 
Hence for hopping along first path to vanish, $\alpha$ and $\beta$ can be chosen such that $\left((-1)^{n} e^{-i(\Ac_{x,y} - \Ac_{x+1,y})\frac{T}{2}} -1 \right) = 0 $ or $\left((-1)^{n} e^{-i(\Ac_{x+1,y} - \Ac_{x+2,y})\frac{T}{2}} -1 \right) =0 $.
\eq{ \nonumber
\Rightarrow \alpha \frac{T}{2} = n \pi ~~\text{or}~~ \beta \frac{T}{2} = n \pi,\\ \nonumber
\Rightarrow \alpha = n \omega ~~\text{or}~~\beta = n \omega
}
For the second path, $\Ac_{x,y} - \Ac_{x+1,y} = \alpha$ and $\Ac_{x+1,y} - \Ac_{x+1,y+1} = \alpha$ for $x$ odd and $\Ac_{x,y} - \Ac_{x+1,y} = \beta$ and $\Ac_{x+1,y} - \Ac_{x+1,y+1} = \beta$ for $x$ even. 
Going through the same steps as that of the first path, for chiral symmetry breaking diagonal hopping to vanish $\left((-1)^{n} e^{-i(\Ac_{x,y} - \Ac_{x+1,y})\frac{T}{2}} -1 \right) = 0 $ or $\left((-1)^{n} e^{-i(\Ac_{x+1,y} - \Ac_{x+1,y+1})\frac{T}{2}} -1 \right) =0 $ for both odd and even $x$.
\eq{\nonumber
\Rightarrow \alpha \frac{T}{2} = n \pi ~~\text{and}~~ \beta \frac{T}{2} = n \pi,\\ \nonumber
\Rightarrow \alpha = n \omega ~~\text{and}~~\beta = n \omega
}
Hence for the chiral symmetry breaking terms to vanish $\alpha = \beta = n \omega$ or $\alpha + \beta = 2 n \omega$. 
Note that this condition is the same as the one chosen for restoration of sublattice symmetry in the zeroth order effective time-independent Hamiltonian as explained in Sec.~\ref{highfrequency}.
If we follow the same steps for second and third terms in Eq.~\eqref{Appendixrotating} to calculate the rest of the first order terms then also we will arrive at the same conclusion. 
Thus it is possible for us to tune the $\alpha$ and $\beta$ in such a way so that the second neighbour hopping emerging from the first order terms in BW perturbation theory goes to zero thus maintaining the chiral symmetry of the system. 
Hence we can conclude that if our original time dependent model has only nearest neighbour hopping then the effective chiral symmetric time-independent Hamiltonian also contains  nearest neighbour hopping terms only.
Generalising this idea, for our model the range of hopping of effective time-independent Hamiltonian will be the same as that of the original time dependent Hamiltonian.\\
To compare results from BW perturbation theory, we also calculate the first order term in Floquet-Magnus expansion given by \cite{Eckardt},
\eq{\label{floquetfirstorder}
\Hc^{F}_{2} = \sum_{n \neq 0} \frac{1}{n \omega}\left( \mathit{H}_{m} \mathit{H_{-m}} + e^{im\omega t_{0}} [\mathit{H_{0}},\mathit{H_{m}}] \right)
}
where, $\mathit{H}_{-m}$, $\mathit{H}_{m}$ and $\mathit{H}_{0}$
are the fourier components of $\Hc(t)$~\eqref{Appendix2dexact}.~The $n^{th}$ fourier component of $\Hc(t)$ is $H_{n}{=} \frac{1}{T} \int_{0}^{T} \Hc(t) e^{i n \omega t} dt $.
As observed from the above BW perturbation theory calculations, for chiral symmetry preserving hopping $\frac{1}{n \omega} \mathit{H}_{m} \mathit{H_{-m}} = 0$ and $\alpha = \beta = n \omega$. 
The commutator $[H_{0},H_{m}]$ vanishes as shown below.
The fourier components of $\Hc(t)$ \eqref{Appendix2dexact}  are (considered nearest neighbour hopping term for calculations),
\eq{\nonumber
H_{0} &= \frac{iJ_{x,y;x+1,y}\omega}{\pi(\Ac_{x,y}-\Ac_{x+1,y})}\left(e^{-i\frac{(\Ac_{x,y}-\Ac_{x+1,y})T}{2}} -1\right) c^{\dagger}_{x,y}c_{x+1,y},\\ \label{fouriercomponents}
H_{m} &= \frac{i J_{x',y',x'+1,y'}\omega}{2\pi}\left((-1)^{m} e^{-i(\Ac_{x',y'} - \Ac_{x'+1,y'})\frac{T}{2}} -1 \right)\left(\frac{1}{(\Ac_{x',y'} - \Ac_{x'+1,y'})-m \omega} + \frac{1}{(\Ac_{x',y'} - \Ac_{x'+1,y'})+m \omega}\right)c^{\dagger}_{x',y'}c_{x'+1,y'}
}
Substituting Eq.~\eqref{fouriercomponents} in the second term of Eq.~\eqref{floquetfirstorder} gives a commutator as shown below.
\eq{\nonumber
[c^{\dagger}_{x,y}c_{x+1,y}, c^{\dagger}_{x',y'}c_{x'+1,y'}] = \delta_{x+1,x'}\delta_{y,y'} c^{\dagger}_{x,y}c_{x'+1,y'} - \delta_{x,x'+1}\delta_{y,y'} c^{\dagger}_{x',y'}c_{x+1,y}
}
Just like first order term in BW perturbation theory here also we get two different path for hopping.
Using the same arguments as in BW perturbation theory, here also for the chiral symmetry to be preserved hopping along both paths should vanish ($H_{m}=0$) and $\alpha = \beta = n \omega$. 
The conclusion remains the same even when the hopping are long-range.
Hence according to Floquet- Magnus expansion also only zeroth order term survives in the effective chiral symmetric time- independent Hamiltonian,
\eq{\nonumber
\Hc_{eff} &= H_{0} =  \Hc^{0}_{BW}
}
\eq{\nonumber 
\Hc_{eff}  =  \sum_{y} \sum_{x,x'}\frac{iJ_{xy;x'y}\omega}{\pi(\Ac_{xy}-\Ac_{x'y})}\left(e^{-i\frac{(\Ac_{xy}-\Ac_{x'y})T}{2}} -1\right) c^{\dagger}_{x,y}c_{x',y} +  \sum_{x=1,3,5,..} \sum_{y,y'}\frac{iJ_{xy;xy'}\omega}{\pi(\Ac_{xy}-\Ac_{xy'})}\left(e^{-i\frac{(\Ac_{xy}-\Ac_{xy'})T}{2}} -1\right) c^{\dagger}_{x,y}c_{x,y'} \\ \nonumber-\sum_{x= 2,4,6,..} \sum_{y,y'}\frac{iJ_{xy;xy'}\omega}{\pi(\Ac_{xy}-\Ac_{xy'})}\left(e^{-i\frac{(\Ac_{xy}-\Ac_{xy'})T}{2}} -1\right) c^{\dagger}_{x,y}c_{x,y'}.
}
\section{Comparison of BW theory with exact diagonalization study.}\label{appendixb}
In this section we show that the exact diagonalization of the Floquet Hamiltonian explained in Eq.~\eqref{effectivefloquetHamiltonian} and that of the effective time-independent Hamiltonian obtained from BW perturbation theory with renormalized hopping as in Eq.~\eqref{realspacehopping} gives us the same results in the limit $\omega > J_{xy;x'y'}$~\eqref{modifiedrealqd}.
However the Floquet Hamiltonian \eqref{effectivefloquetHamiltonian} matches accurately with the effective time-independent Hamiltonian \eqref{realspacehopping} only for $\omega >> J_{xy;x'y'}$~\eqref{modifiedrealqd}.
We analyze the energy spectrum of Floquet Hamiltonian explained in Sec.~\ref{dynamics} (the eigenvalues are sorted in ascending order) and the real space energy spectrum of effective time-independent Hamiltonian from BW perturbation theory for different winding numbers as shown in Fig.~\ref{A1} and Fig.~\ref{A2}.
\begin{figure}[h]
\includegraphics[width=0.5\textwidth]{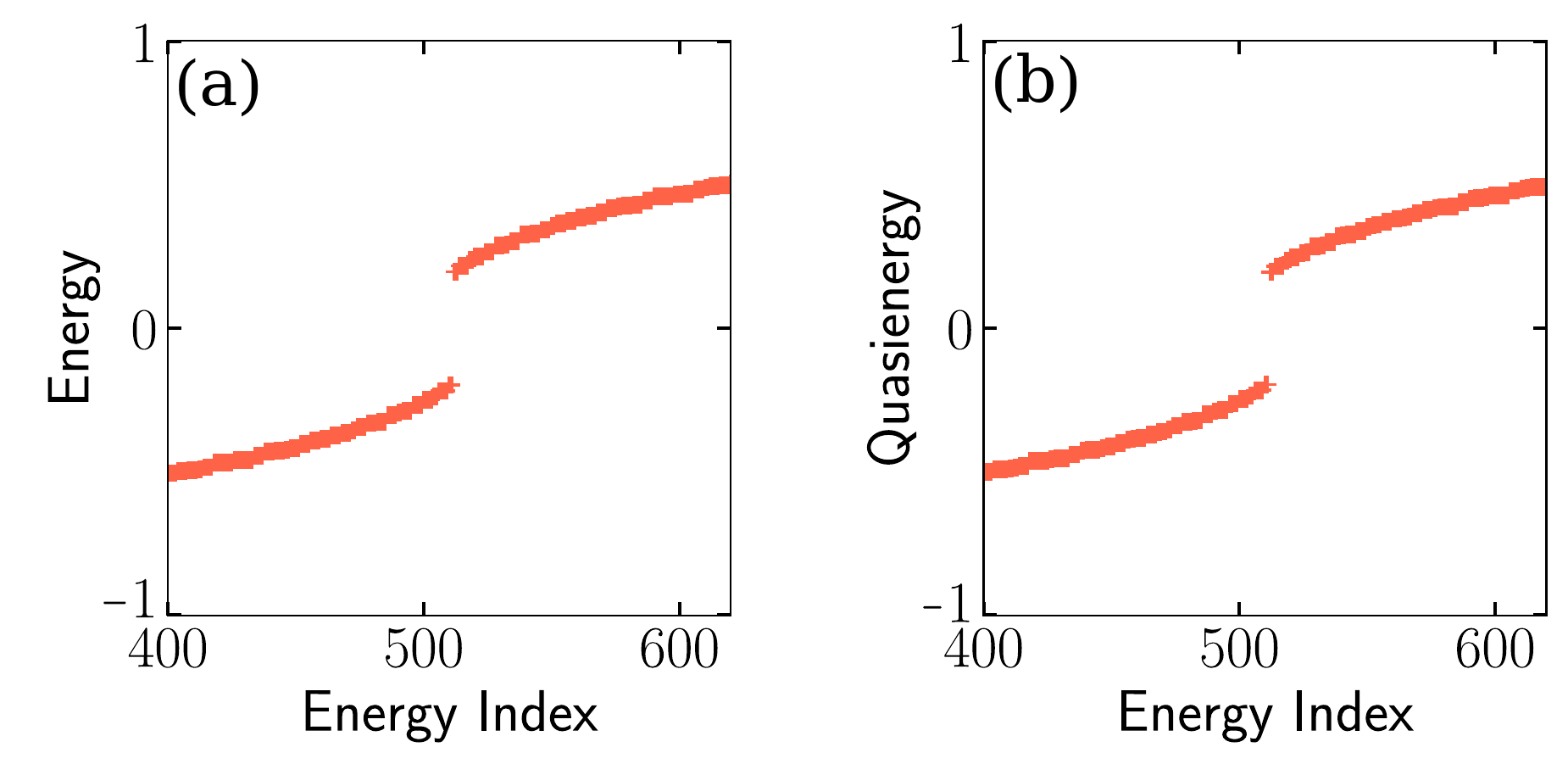}
\caption{(a) Shows the real space energy spectrum of effective Hamiltonian \eqref{realspacehopping}.~The parameter values are, $\alpha/\omega = 0.8$, $r =1$, $\lambda = 1.5$ and $\omega = 1$. Here $\nu_{2D} = 0$ with no zero energy corner states.
(b) Shows the quasienergy spectrum of Floquet Hamiltonian~\eqref{effectivefloquetHamiltonian} for same set of parameters but with $\omega = 10$ and $\alpha = 8$ such that $\alpha/\omega = 0.8$. Here also there are no zero quasienergy corner states.}\label{A1}
\end{figure}
\begin{figure}[h]
\includegraphics[width=0.45\columnwidth]{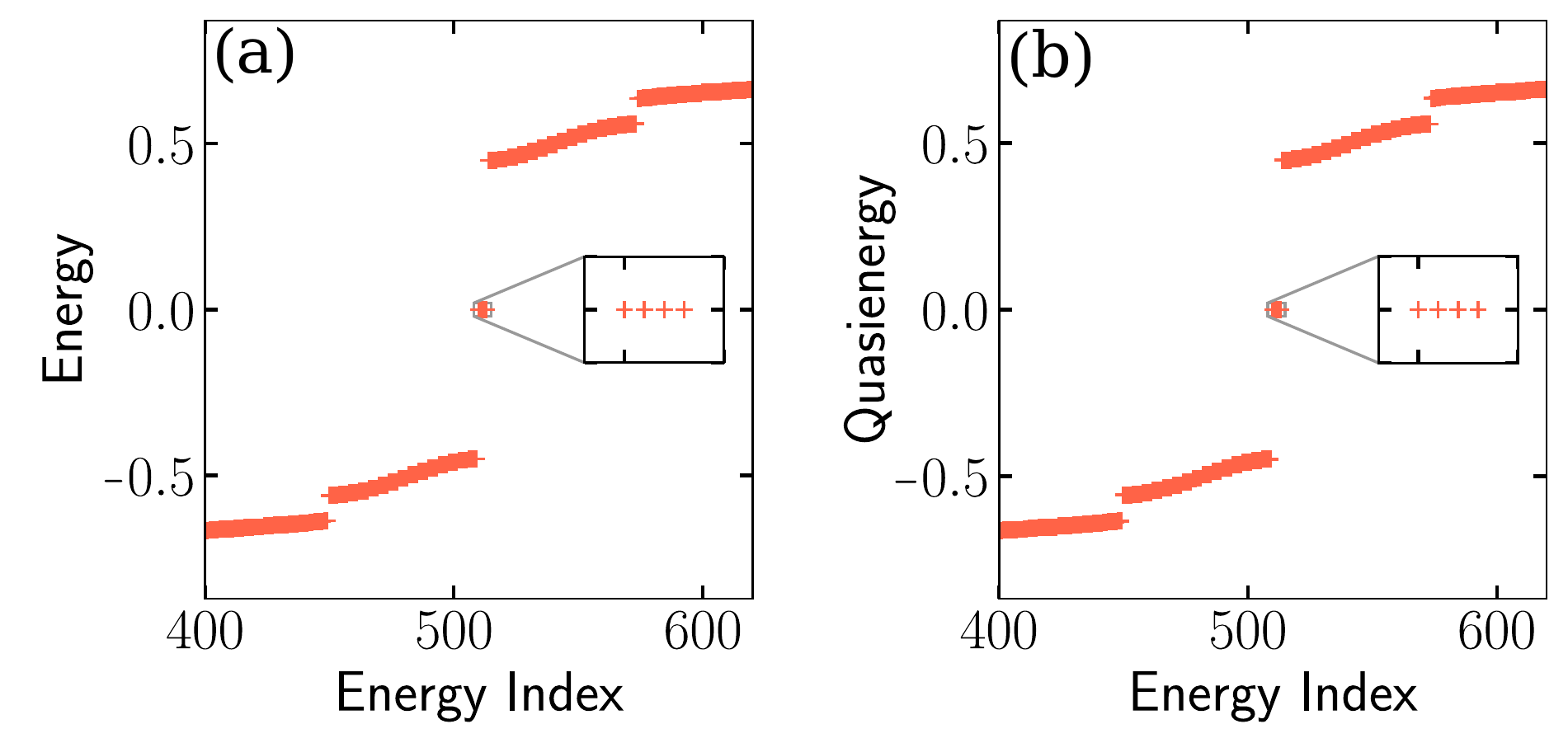}
\includegraphics[width=0.45\columnwidth]{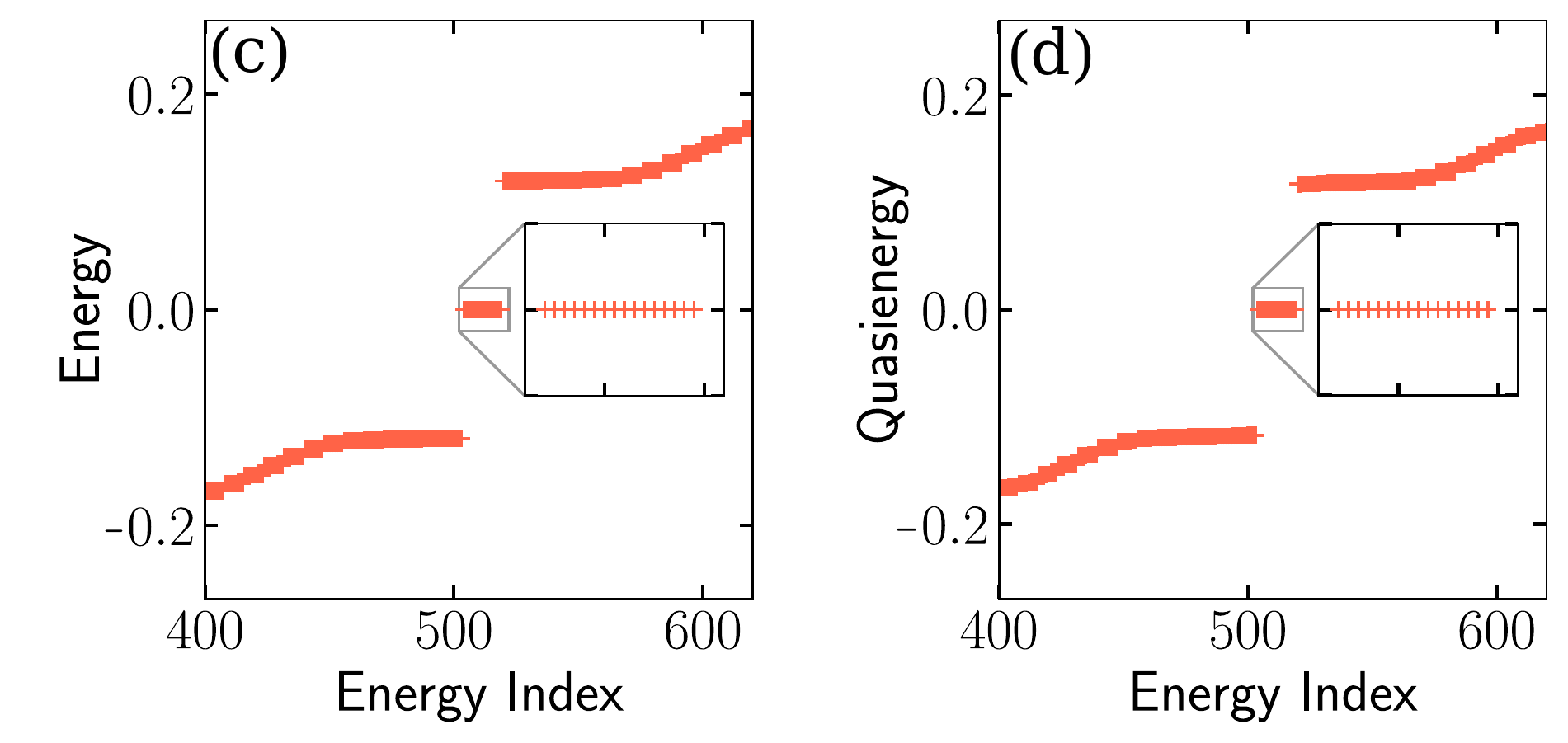}
\caption{(a) The real space spectrum of effective Hamiltonian~\eqref{realspacehopping}.~ The parameter values are, $r =1$,~$\alpha/\omega = 1.8$, $\lambda = 1.5$ and $\omega = 1$.~Here $\nu_{2D} = 1$ and  number of corner states is four.
(b) Shows the  quasienergy spectrum of Floquet Hamiltonian~\eqref{effectivefloquetHamiltonian} for same set of parameters but with $\alpha = 18$ and $\omega = 10$ such that $\alpha/\omega = 1.8$.~Here we get four fold degenerate corner states at zero quasienergy.
(c) The real space spectrum of Hamiltonian~\eqref{realspacehopping} for $\nu_{2D} = 2$ with $r =3$, $\alpha/\omega = 4.1$ $\lambda = 1.5$ and $\omega = 1$.
For Floquet Hamiltonian we take $\alpha = 41$ and $\omega = 10$ such that $\alpha/\omega = 4.1$ and the corresponding quasienergy spectrum is shown in (d).~In both (c) and (d) we get $16$ zero energy (quasienergy) modes localized at the corners of the lattice.}\label{A2}
\end{figure}
While diagonalizing Eq.~\eqref{effectivefloquetHamiltonian} we take the number of unit cells to be two times the number taken while diagonalizing the effective Hamiltonian with renormalized hopping as in Eq.~\eqref{realspacehopping}.
This is because the sublattice structure evolves naturally while studying the exact dynamics using Floquet Hamiltonian.
From Fig.~\ref{A1} it is clear that the energy spectrum of Floquet Hamiltonian converges with that of time-independent Hamiltonian from BW perturbation theory for values of $\omega$ which are much greater than the hopping amplitude ($J_{xy;x'y'}$) of the original time dependent model~\eqref{modifiedrealqd}. 
Here we have taken $\omega \gtrsim 20 J_{xy;x'y'}$. 
Fig.~\ref{A1} and Fig.~\ref{A2} corresponds to three different set of two dimensional winding number. 
As we can see from the figures, the quasienergy  spectrum is similar to the real energy spectrum and we get same number of zero quasienergy states as that of the zero energy states of real space energy spectrum of effective Hamiltonian~\eqref{realspacehopping}.
The Floquet states corresponding to these zero quasienergy modes are also found to be localized at the corners of the lattice. This also confirms the validity of Brillouin-Wigner perturbation theory in the high-frequency regime.
\end{widetext}
\end{document}